\pdfoutput=1
\documentclass[aps,prd,showpacs,superscriptaddress,amsmath,amsfont,amssymb,graphicx]{revtex4}
\usepackage{color,graphicx,epsfig} 
\usepackage{stmaryrd}
\usepackage{ifpdf}
\usepackage{amsmath} 
\usepackage{amssymb} 
\usepackage{bm}
\usepackage{color}
\usepackage{braket}
\usepackage[english]{babel}

\newcommand{\be}{\begin{eqnarray}}
\newcommand{\ee}{\end{eqnarray}}
\newcommand{\op}{{\mathcal O}}
\newcommand{\fig}[1]{Figure \ref{#1}}
\newcommand{\epsUV}{\epsilon_{\mathrm{UV}}}
\newcommand{\epsIR}{\epsilon_{\mathrm{IR}}}
\newcommand{\ii}{\mathrm{i}}

\newcommand{\uc}{\hat{y}}
\newcommand{\xc}{\hat{x}}
\newcommand{\Brfrac}{\mathrm{Br}^{\mathrm{NLO}}/\mathrm{Br}^{\mathrm{LO}}}

\definecolor{Red}{rgb}{1.,0.,0.}

\begin{document}
\title{QCD Corrections to Flavor Changing Neutral Coupling  Mediated Rare Top Quark Decays}

\author{Jure Drobnak} 
\email[Electronic address:]{jure.drobnak@ijs.si} 
\affiliation{J. Stefan Institute, Jamova 39, P. O. Box 3000, 1001
  Ljubljana, Slovenia}

\author{Svjetlana Fajfer} 
\email[Electronic address:]{svjetlana.fajfer@ijs.si} 
\affiliation{J. Stefan Institute, Jamova 39, P. O. Box 3000, 1001
  Ljubljana, Slovenia}
\affiliation{Department of Physics,
  University of Ljubljana, Jadranska 19, 1000 Ljubljana, Slovenia}

\author{Jernej F. Kamenik}
\email[Electronic address:]{jernej.kamenik@ijs.si} 
\affiliation{J. Stefan Institute, Jamova 39, P. O. Box 3000, 1001
  Ljubljana, Slovenia}

\date{\today}

\begin{abstract}
Recently we have presented an analysis of flavor changing neutral coupling mediated radiative top quark decays at next-to-leading order in QCD. In the present paper we provide the details of the calculation of QCD corrections to $t\to q \gamma$ and $t\to q Z$ decays within the effective theory approach including operator mixing.  In particular, we calculate virtual matrix element corrections and the corresponding bremsstrahlung contributions. In the case of $t\to q \gamma$ we study the effects of kinematic cuts on the extracted branching ratios. Analytical formulae are given at all stages of the calculation. We find that the $t \to q \gamma$ decay can be used to probe also the effective operators mediating $t \to q g$ processes, since these can naturally contribute $10\%$ or more to the radiative decay, given typical experimental cuts on the decay kinematics at hadron colliders. Conversely, we argue that any positive experimental signal of the $t \to q g$ process would indicate a natural lower bound on $t \to q \gamma$ decay rate.
\end{abstract}


\maketitle
\section{Introduction}
The standard model (SM) predicts highly suppressed flavor changing neutral current (FCNC) processes of the top quark  ($t\to q V,\,V=Z,\gamma,g,\, q=c,u$) while new physics beyond the SM (NP) in many cases lifts this suppression (for a recent review c.f.~\cite{AguilarSaavedra:2004wm}). It has been pointed out recently, that top quark FCNC phenomenology is crucial in constraining a wide class of NP scenarios, where new flavor structures are present but can be aligned with the SM Yukawas in the down sector~\cite{Fox:2007in, Gedalia:2010zs, Gedalia:2010mf, Datta:2009zb}. Top quark FCNCs can be probed both in production and in decays. Presently the most stringent bound on the $\mathrm{Br}(t\to q Z)$ comes from a search performed by the CDF collaboration $\mathrm{Br}(t\to q Z) < 3.7\%$ at $95\%$ C.L. \cite{:2008aaa}. The photonic decay is presently mostly constrained by the ZEUS collaboration $\mathrm{Br}(t\to q \gamma) < 0.59 \%$ at $95\%$ C.L.~\cite{Chekanov:2003yt}. On the other hand the most stringent present limit on $\mathrm{Br}(t\to q g)$ comes from the single top production total cross-section measurement of CDF and yields $\mathrm{Br}(t \to u g) < 0.039\%$ and $\mathrm{Br}(t \to c g) < 0.57\%$ at $95\%$ C.L.~\cite{Aaltonen:2008qr}. The LHC will be producing about $80,000$ $t\bar t$ events per day at the luminosity $L = 10^{33} \mathrm{cm}^{-2}\mathrm{s}^{-1}$ and will be able to access rare top decay branching ratios at the $10^{-5}$ level with $10 \mathrm{fb}^{-1}$~\cite{Carvalho:2007yi}.

Recently~\cite{Drobnak:2010wh,Zhang:2008yn} the $t\to q V$ decays mediated by effective FCNC couplings have been investigated at next-to-leading order (NLO) in QCD. In~\cite{Zhang:2008yn} it was found that $t\to q g$ receives almost $20\%$ enhancement while corrections to the $t\to c \gamma, Z$ branching ratios are much smaller.  Contributions of additional operators in $t\to q Z$ and the effects of operator mixing induced by QCD corrections have been identified in~\cite{Drobnak:2010wh}.  In the case of $t\to q \gamma$ decay in particular the QCD corrections generate a nontrivial photon spectrum and the correct process under study is actually $t\to q \gamma g$. Experimental signal selection for this mode is usually based on kinematical cuts, significantly affecting the extracted bounds on the effective FCNC couplings. The main implications of this observation were presented in~\cite{Drobnak:2010wh}, while the details of the underlying calculation and full analytic results are the subject of the present paper. 

This paper is structured as follows. We begin by giving the set of FCNC operators considered and explain our notation. Next we conduct a full study of QCD corrections to $t\to q Z,\gamma$ decays. Contributions from gluonic dipole operators are also taken into account. We present results for the virtual matrix element corrections as well as the corresponding bremsstrahlung rates. In the $t\to q \gamma$ channel we also study the relevance of kinematical cuts on the photon energy and the angle between the photon and the jet stemming from the final state quark. We present our results in analytical form and also give numerical values to estimate the significance of NLO contributions.


\section{Framework}
In writing the effective top FCNC Lagrangian we rely on the notation of ref.~\cite{AguilarSaavedra:2004wm, AguilarSaavedra:2008zc}. Hermitian conjugate and chirality flipped operators are implicitly contained in the Lagrangian and contributing to the relevant decay modes
\be
{\mathcal L}_{\mathrm{eff}} = \frac{v^2}{\Lambda^2}a_L^{Z}\op_{L}^Z
+\frac{v}{\Lambda^2}\Big[b^{Z}_{LR}\op_{LR}^{Z}+b^{\gamma}_{LR}\op_{LR}^{g}+b^{g}_{LR}\op_{LR}^{\gamma}
\Big] + (L \leftrightarrow R) + \mathrm{h.c.}\,.
\label{eq:Lagr}
\ee
To explain the notation, operators considered are
\begin{align}
\op^{Z}_{L,R} &= g_Z Z_{\mu}\Big[\bar{q}_{L,R}\gamma^{\mu}t_{L,R}\Big]\,, &
\op^{Z}_{LR,RL} &= g_Z Z_{\mu\nu}\Big[\bar{q}_{L,R}\sigma^{\mu\nu}t_{R,L}\Big]\,,\nonumber\\
\op^{\gamma}_{LR,RL} &= e A_{\mu\nu}\Big[\bar{q}_{L,R}\sigma^{\mu\nu}t_{R,L}\Big]\,, &
\op^{g}_{LR,RL} &= g_s G^a_{\mu\nu}\Big[\bar{q}_{L,R}\sigma^{\mu\nu}T_a t_{R,L}\Big]\,,
\label{ops}
\end{align}
where $q_{R,L} = (1\pm\gamma_5)q/2$, $\sigma_{\mu\nu} = i[\gamma_\mu,\gamma_\nu]/2$ and $g_Z = 2 e/\sin 2 \theta_W$. Here
$\theta_W$ stands for the Weinberg angle, while $e=\sqrt{4\pi \alpha_{}}$ and $g_s=\sqrt{4\pi \alpha_s}$. Furthermore $V(A,Z)_{\mu\nu} = \partial_{\mu} V_\nu - \partial_\nu V_\mu$ and $G^a_{\mu\nu} =  \partial_{\mu} G^a_\nu - \partial_\nu G^a_\mu + g f_{abc} G_\mu^b G_\nu^c $ where and $T^a$ and $f_{abc}$ are the $SU(3)$ color group generators and structure constants respectively. Finally $v=246$~GeV is the electroweak condensate and $\Lambda$ is the effective scale of NP. In the remainder of the paper, since there is no mixing between chirality flipped operators we shorten the notation, setting  $a$ and $b$ to stand for $a_L$, $b_{LR}$ or $a_R$, $b_{RL}$. 

Note that in principle, additional, four-fermion operators might be induced at the high scale which will also give contributions to $t\to q V$ processes, however these are necessarily $\alpha_s$ suppressed. On the other hand, such contributions can be more directly constrained via e.g. single top production measurements and we neglect their effects in the present study. We also assume the effective $a,b$ couplings are defined near the top quark mass scale at which we evaluate virtual matrix element corrections and $\alpha_s$. A translation to a higher scale matching is governed by the anomalous dimensions of the effective operators and can be performed consistently using RGE methods. The procedure and its implications has been discussed in ref.~\cite{Drobnak:2010wh} and we do not repeat it here. In our calculation we neglect the mass of the final state $(c,u)$ quark and regulate UV as well as IR divergences by working in $d=4+\epsilon$ dimensions.

\section{$t\to q Z$ decay}
The total $t\to q Z$ decay width, mediated by operators in Lagrangian~(\ref{eq:Lagr}) and including leading QCD corrections can be written in the following form
\be
\Gamma^Z &=& |a^Z|^2\frac{v^4}{\Lambda^4} \Gamma_{a}^Z + \frac{v^3m_t}{\Lambda^4} \left[2\mathrm{Re}\{a^{Z*}b^g\} \Gamma^Z_{ag} + 2\mathrm{Re}\{b^{Z*}a^Z\} \Gamma^Z_{ab}
 - 2\mathrm{Im}\{a^{Z*}b^g\} \tilde{\Gamma}^Z_{ag}\right]\label{dw}\\
&&\hspace{1.81cm}+ \frac{v^2 m_t^2}{\Lambda^4}\left[ |b^Z|^2 \Gamma^Z_{b} 
+|b^g|^2 \Gamma^Z_{g} +2\mathrm{Re}\{b^{Z*}b^g\} \Gamma^Z_{bg} -2\mathrm{Im}\{b^{Z*}b^g\}\tilde{\Gamma}^Z_{bg} \right]\nonumber\,. 
\ee
It includes tree level contributions, one loop virtual corrections and gluon bremsstrahlung processes (actually $t\to q Z g$) which we will analyze separately in the following two subsections. In the end of the section we present numerical results of the combined effects.

\subsection{Tree level expressions}
At the tree level we only have $\Gamma_{a}^Z$, $\Gamma_b^Z$ and $\Gamma_{ab}^Z$ contributions, which we write in $4+\epsilon$ dimensions as 
\be
\Gamma_{a}^{Z(0)}&=&\lim_{\epsilon\to 0}\frac{m_t}{16\pi}g_Z^2(1-r_Z)^2 \Gamma(1+\frac{\epsilon}{2})(1-r_Z)^{\epsilon}\frac{1}{2r_Z}\big(1+(2+\epsilon)r_Z\big)\,,\\
\Gamma_{b}^{Z(0)}&=&\lim_{\epsilon\to 0}\frac{m_t}{16\pi}g_Z^2(1-r_Z)^2 \Gamma(1+\frac{\epsilon}{2})(1-r_Z)^{\epsilon} 2(2+\epsilon+r_Z)\,,\nonumber\\
\Gamma_{ab}^{Z(0)}&=&\lim_{\epsilon\to 0}\frac{m_t}{16\pi}g_Z^2(1-r_Z)^2 \Gamma(1+\frac{\epsilon}{2})(1-r_Z)^{\epsilon}(3+\epsilon)\,,\nonumber
\ee
where $r_Z=m_Z^2/m_t^2$.

\subsection{Virtual corrections}
\begin{figure}[h]
\begin{center}
\includegraphics[scale=0.8]{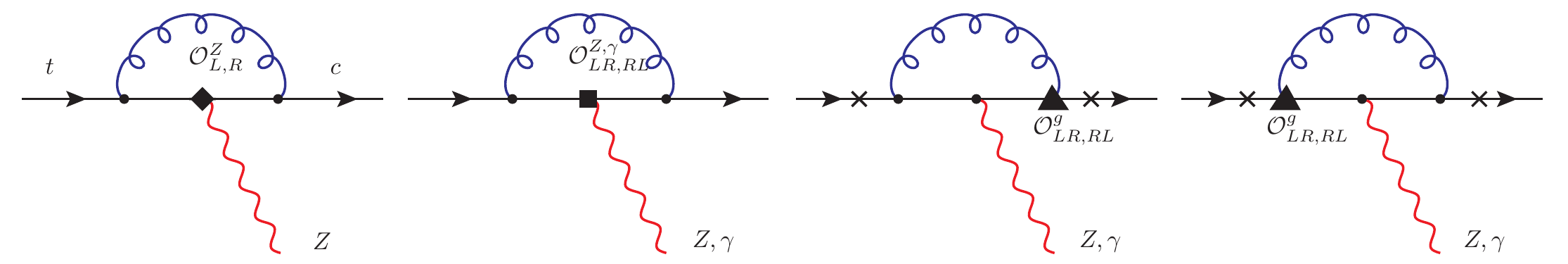}
\end{center}
\caption{One loop virtual corrections to $t\to cZ,\gamma$ decay. Crosses mark additional points
from which the $Z,\gamma$ boson can be emitted.}
\label{virtcorrections}
\end{figure}

At the one loop level, $t\to q Z,\gamma$ decay rates receive contributions from Feynman diagrams in  \fig{virtcorrections}. Besides the one loop gluon corrections to the $Z,\gamma$ FCNC operator matrix elements, we also include contributions from the gluonic FCNC dipole operators. The later first appear at $\mathcal O(\alpha_s)$ with the emission of the $Z,\gamma$ proceeding through SM couplings.
On the other hand, one particle reducible diagrams with gluon corrections attached to the external legs are not presented in  \fig{virtcorrections}. They are taken into account by proper quark field renormalization $q\to \sqrt{Z_q}q$. Since the final state light quark is considered to be massless, the corresponding field renormalization differs from the one of the initial top quark. Using the on-shell renormalization conditions we get
\be
Z_t &=& 1+ 
\frac{\alpha_s}{4\pi}C_F \frac{\Gamma(1-\frac{\epsilon}{2})}{(4\pi)^{\epsilon/2}}\left(\frac{m_t}{\mu}\right)^{\epsilon}
\Big[\frac{2}{\epsUV}+\frac{4}{\epsIR}-4\Big]\,,\\
Z_q &=& 1+ 
\frac{\alpha_s}{4\pi}C_F \frac{\Gamma(1-\frac{\epsilon}{2})}{(4\pi)^{\epsilon/2}}\left(\frac{m_t}{\mu}\right)^{\epsilon}
\Big[\frac{2}{\epsUV}-\frac{2}{\epsIR}\Big]\,,\nonumber
\ee
where $\mu$ is the renormalization scale parameter and $C_F=4/3$. The resulting $t\to qZ$ decay amplitude including one loop virtual effects can be written in terms of six form factors
\be
A_{t\to qZ} &=& \Big[\frac{v^2}{\Lambda^2} a^Z \big(1+\frac{\alpha_s}{4\pi}C_F F^Z_a\big)
  + \frac{v}{\Lambda^2}b^Z \frac{\alpha_s}{4\pi}C_F F_{ab}^Z 
  + \frac{v}{\Lambda^2}b^g \frac{\alpha_s}{4\pi}C_F F_{ag}^Z\Big]\langle \op_{L,R}^Z\rangle\\
&+& \Big[\frac{v}{\Lambda^2} b^Z \big(1+\frac{\alpha_s}{4\pi}C_F F^Z_b\big)
  + \frac{v^2}{\Lambda^2}a^Z \frac{\alpha_s}{4\pi}C_F F_{ba}^Z 
  + \frac{v}{\Lambda^2}b^g \frac{\alpha_s}{4\pi}C_F F_{bg}^Z\Big]\langle \op_{LR,RL}^Z\rangle\,.\nonumber
\ee
The form factors are
\begin{subequations}\label{ZForms}
\be
F_{a}^Z&=&C_{\epsilon}\left[
-\frac{4}{\epsIR^2}+\frac{5-4\log(1-r_Z)}{\epsIR}-2\log^2(1-r_Z)+3\log(1-r_Z)-2\mathrm{Li}_2(r_Z)-6\right]\,,\label{Fa}\\
F_{b}^Z&=&C_{\epsilon}
\Big[
-\frac{4}{\epsIR^2}+\frac{5-4\log(1-r_Z)}{\epsIR}+\frac{2}{\epsUV} -2 \log^2(1-r_Z)+4\log(1-r_Z)-2\mathrm{Li}_2(r_Z)-6 \Big]+\delta_{b}^Z\,,\label{Fb}\\
F_{ab}^Z&=&-4 m_t \log(1-r_Z)\,,\label{Fab}\\
F_{ba}^Z&=&-\frac{1}{m_t}\frac{1}{2r_Z}\log(1-r_Z)\,,\label{Fba}
\ee
\end{subequations}
\begin{subequations}\label{glueForms}
\be
F_{ag}^Z&=&m_t\left[\hat v+\hat a+ (\hat v-\hat a)\Big\llbracket\frac{r_Z(4-r_Z)(1+r_Z)}{(1-r_Z)^3}f_1 -\frac{2r_Z(4-r_Z)}{(1-r_Z)^4}f_2-\frac{1-7r_Z+3r_Z^2}{(1-r_Z)^2}+\frac{2 r_Z}{1-r_Z}\log r_Z\Big\rrbracket\right]\,,\label{Fag}\\
F_{bg}^Z&=&C_{\epsilon}\bigg[
2 \hat v \frac{2}{\epsUV}+(\hat v+\hat a)(f_1-2)\label{Fbg}\\
&+&(\hat v-\hat a)\Big\llbracket-\frac{r_Z}{1-r_Z}\log(r_Z) -\ii  \pi- \frac{7/2-4r_Z+2r_Z^2}{(1-r_Z)^2} -\frac{(1+r_Z)(2+r_Z)}{2(1-r_Z)^3}f_1+\frac{2+r_Z}{(1-r_Z)^4}f_2\Big\rrbracket\bigg]+ \delta_{bg}^Z \,.\nonumber
\ee
\end{subequations}
We have defined auxiliary functions $f_1$ and $f_2$ for shorter notation
\be
f_1&=&2\sqrt{\frac{4-r_Z}{r_Z}}\arctan\Big(\sqrt{\frac{r_Z}{4-r_Z}}\Big)\,,\\
f_2&=&-2\mathrm{Li}_2(r_Z-1)+2\arctan\Big(\frac{1-r_Z}{3-r_Z}\sqrt{\frac{4-r_Z}{r_Z}}\Big)
\arctan\Big(\frac{r_Z}{2-r_Z}\sqrt{\frac{4-r_Z}{r_Z}}\Big) \nonumber\\
 &+&2\mathrm{Re}\Big\{\mathrm{Li}_2\Big((1-r_Z)^2
\big(1-\frac{r_Z}{2}\frac{2-r_Z}{1-r_Z}(1+\ii\sqrt{\frac{4-r_Z}{r_Z}})\big)\Big)
-\mathrm{Li}_2\Big(\frac{1-r_Z}{2}\big(2-r_Z-\ii \sqrt{(4-r_Z)r_Z}\big)\Big)\Big\}\,.\nonumber
\ee
$F_a^Z$ and $F_b^Z$ include the quark field renormalization. To rid all the UV divergences, operator renormalization is necessary. This leads to the appearance of counter terms $\delta_b^{Z}$ and $\delta_{bg}^Z$  which are renormalized in the appropriate matching condition in the UV. The two operator renormalization counter terms are evaluated in $\overline{\text{MS}}$ scheme
\be
Z_{b}^Z  &=& 1+\frac{\alpha_s}{4\pi}C_F\delta_{b}^Z\,,\hspace{0.5cm}
\delta_{b}^Z= - \Big(\frac{2}{\epsUV}\ +\gamma-\log(4\pi)\Big)\,,\label{renZ}\\
Z_{bg}^Z &=& 1+\frac{\alpha_s}{4\pi}C_F\delta_{bg}^Z\,,\hspace{0.5cm}
\delta_{bg}^Z= -2 v \Big(\frac{2}{\epsUV} +\gamma-\log(4\pi)\Big)\,.
\ee

On the other hand, $C_{\epsilon}=(m_t/\mu)^{\epsilon}\Gamma(1-\epsilon/2)/(4\pi)^{\epsilon/2}$ is an IR renormalization factor multiplying divergent form factors. Finally, SM $Z$ boson couplings to fermions are defined as $\hat v=T_3 -2\sin\theta_W Q $ and $\hat a = T_3$, where for the up-type quarks $T_3=1/2$ and $Q=2/3$. They appear only in form factors generated by the gluonic dipole operator. 

We were able to crosscheck our expressions for form factors with those found in the literature. Namely, (\ref{ZForms}) agree with the corresponding expressions given in ref. \cite{Ghinculov:2002pe} for the $B\to X_s l^+ l^-$ decay mediated by a virtual photon after taking into account that the dipole operator in \cite{Ghinculov:2002pe} includes a mass parameter which necessitates additional mass renormalization. To some extent we were also able to crosscheck the gluon operator induced form factors (\ref{glueForms}). Namely, we find numerical agreement of the form factor vector component (i.e. setting $\hat a=0$) with the corresponding expressions given in ref. \cite{Ghinculov:2003qd}. The crosscheck is of course only possible in the vector part, since the SM photon coupling appearing in \cite{Ghinculov:2003qd} has no axial component.

After UV renormalization we are still left with IR divergences. These will be canceled at the level of the decay width by the IR divergences associated with the bremsstrahlung process $t\to q Z g$. In order to demonstrate this cancellation explicitly, we write down the contribution of the tree level and virtual corrections to decay widths defined in eq.~(\ref{dw}). The IR divergent parts are
\begin{subequations}\label{IRvirt}
\be
\Gamma_{a}^{Z,\mathrm{virt.}}&=&\Gamma_{a}^{Z(0)}\bigg[1+\frac{\alpha_s}{4\pi} C_F\Big\llbracket-\frac{8}{\epsIR^2}+\frac{-16 \log(1-r_Z)+\frac{4}{1+2 r_Z}+6}{\epsIR}\label{virtIR1} \\
&-& 16 \log^2(1-r_Z)+\frac{2(5+8r_Z)}{1+2 r_Z}\log(1-r_Z)-\frac{\pi^2}{3}-\frac{2(6+7r_Z)}{1+2r_Z} - 4\mathrm{Li}_2(r_Z)\Big\rrbracket\bigg]\,,\nonumber\\
\Gamma_{b}^{Z,\mathrm{virt.}}&=&\Gamma_{b}^{Z(0)}\bigg[1+\frac{\alpha_s}{4\pi}C_F\Big\llbracket-\frac{8}{\epsIR^2}+\frac{-16 \log(1-r_Z)-\frac{8}{2+ r_Z}+10}{\epsIR} \label{virtIR2}\\
&-& 16 \log^2(1-r_Z)
+\frac{2(4+9r_Z)}{2+r_Z}\log(1-r_Z)-\frac{\pi^2}{3}+2\log\left(\frac{m_t^2}{\mu^2}\right) - \frac{2(7+6r_Z)}{2+r_Z}-4\mathrm{Li}_2(r_Z)\Big\rrbracket\bigg]\,,\nonumber \\
\Gamma_{ab}^{Z,\mathrm{virt.}}&=&\Gamma_{ab}^{Z(0)}\bigg[1+\frac{\alpha_s}{4\pi}C_F\Big\llbracket-\frac{8}{\epsIR^2}+\frac{-16 \log(1-r_Z)+\frac{22}{3}}{\epsIR} \label{virtIR3}\\
&-&16 \log^2(1-r_Z)-\frac{2(2-15 r_Z)}{3 r_Z}\log(1-r_Z)-\frac{\pi^2}{3}-\frac{26}{3}-4\mathrm{Li}_2(r_Z)\Big\rrbracket\bigg]\,.\nonumber
\ee
\end{subequations}
Remaining contributions are induced by the gluonic operator and are IR finite
\begin{subequations}
\be
\Gamma_{ag}^{Z,\mathrm{virt.}}&=&\Gamma_{ab}^{Z(0)}\frac{\alpha_s}{4\pi}C_F \bigg[
2 \hat v \log\left(\frac{m_t^2}{\mu^2}\right)+(\hat v-\hat a)\Big[\frac{1}{3}\log(r_Z)+\frac{2f_2}{3(1-r_Z)^2}\Big]\\
&+&\frac{2}{3}f_1\frac{\hat a(2-r_Z)+\hat v(1-2 r_Z)}{1-r_Z}+\frac{\hat a}{3} (4+\frac{1}{r_Z})-\frac{14\hat v}{3}\bigg]\,,\nonumber\\
\Gamma_{bg}^{Z,\mathrm{virt.}}&=&\Gamma_{b}^{Z(0)}\frac{\alpha_s}{4\pi}C_F\bigg[
2 \hat v \log\left(\frac{m_t^2}{\mu^2}\right) + (\hat v-\hat a)\Big[\frac{r_Z}{2+r_Z}\log(r_Z)+ \frac{4 f_2}{(1-r_Z)^2(2+r_Z)}\Big]\\
&+& f_1\frac{\hat a (4+r_Z-r_Z^2) - \hat v(3+r_Z)r_Z}{(1-r_Z)(2+r_Z)}
- \hat v\frac{11+4r_Z}{2+r_Z}+ \hat a \frac{6}{2+r_Z}\bigg]\,,\nonumber\\
\tilde{\Gamma}_{ag}^{Z,\mathrm{virt.}}&=&\Gamma_{ab}^{(0)} \frac{\alpha_s}{4\pi}C_F (\hat v-\hat a)(-\pi)\,,\\
\tilde{\Gamma}_{bg}^{Z,\mathrm{virt.}}&=&\Gamma_{b}^{(0)}\frac{\alpha_s}{4\pi}C_F (\hat v-\hat a)(-\pi)\,.
\ee
\end{subequations}

\subsection{Bremsstrahlung Contributions}
\begin{figure}[h]
\begin{center}
\includegraphics[scale=0.8]{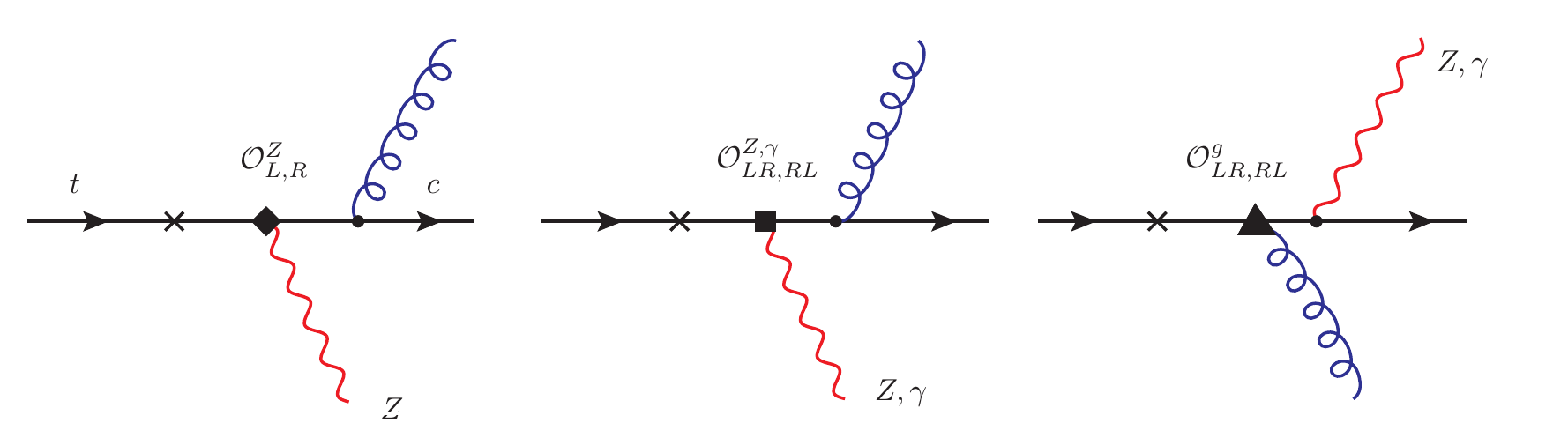}
\end{center}
\caption{Feynman diagrams for $t\to cgZ,\gamma$ bremsstrahlung process. Crosses mark additional points form which the gluon or
$Z,\gamma$ boson can be emitted.}
\label{feynbrems}
\end{figure}

The relevant Feynman diagrams contributing to $t\to q g Z,\gamma$ bremsstrahlung processes are given in Figure \ref{feynbrems}. At the level of the decay width these diagrams give contributions of the same order in $\alpha_s$ as the one-loop virtual corrections presented in the last subsection. Soft and collinear IR divergences emerge in the phase space integration. 
Below we give analytical formulae for the bremsstrahlung contributions to the decay widths in eq.~(\ref{dw}). There are three IR divergent contributions
\begin{subequations}\label{IRbrems}
\be
\Gamma_{a}^{Z,\mathrm{brems.}}&=& \Gamma_a^{Z(0)}\frac{\alpha_s}{4\pi}C_F\bigg[
\frac{8}{\epsIR^2}+\frac{16\log(1-r_Z)-\frac{4}{1+2r_Z}-6}{\epsIR} + 16 \log^2(1-r_Z)- 4\log (r_Z)\log(1-r_Z)\label{bremsIR1}\\
&-& 4\frac{5 + 6 r_Z}{1+2r_Z}\log(1-r_Z)-\frac{4(1-r_Z-2r_Z^2)r_Z}{(1-r_Z)^2(1+2r_Z)} \log(r_Z)- \pi^2
-4\mathrm{Li}_2(r_Z) +\frac{7+r_Z}{(1-r_Z)(1+2r_Z)}+10\bigg]\,,\nonumber\\
\Gamma_{b}^{Z,\mathrm{brems.}}&=& \Gamma_b^{Z(0)}\frac{\alpha_s}{4\pi}C_F\bigg[
\frac{8}{\epsIR^2}+\frac{16\log(1-r_Z)+\frac{8}{2+r_Z}-10}{\epsIR} + 16\log^2(1-r_Z)-4\log(r_Z)\log(1-r_Z)\label{bremsIR2}\\
&-&4\frac{6+5r_Z}{2+r_Z}\log(1-r_Z)-\frac{4(2-2r_Z-r_Z^2)r_Z}{(1-r_Z)^2(2+r_Z)}\log(r_Z)
-\pi^2-4\mathrm{Li}_2(r_Z)-\frac{4-8r_Z}{(1-r_Z)(2+r_Z)}+\frac{43}{3}\bigg]\,,\nonumber\\
\Gamma_{ab}^{Z,\mathrm{brems.}}&=& \Gamma_{ab}^{Z(0)}\frac{\alpha_s}{4\pi}C_F\bigg[
\frac{8}{\epsIR^2}+\frac{16\log(1-r_Z)-\frac{22}{3}}{\epsIR} + 16\log^2(1-r_Z)-4\log(r_Z)\log(1-r_Z)\label{bremsIR3}\\
&-&\frac{44}{3}\log(1-r_Z)-\frac{4(3-2r_Z)r_Z}{3(1-r_Z)^2}\log(r_Z)
-\pi^2-4\mathrm{Li}_2(r_Z)-\frac{4}{3(1-r_Z)}+\frac{47}{3}\bigg]\,.\nonumber
\ee
\end{subequations}
Summing these decay widths with the one-loop virtually corrected $t\to q Z$ rates into $\Gamma_{X}^{Z} = \Gamma_{X}^{Z,\mathrm{virt.}}+\Gamma_{X}^{Z,\mathrm{brems.}}$ for $X=a,\, b,\, ab$, we obtain combined IR finite expressions 
\begin{subequations}\label{GjustZ}
\be
\Gamma_a^{Z}&=&\Gamma_{a}^{Z(0)}\Bigg[1+\frac{\alpha_s}{4\pi}C_F\bigg[
-4\log(1-r_Z)\log(r_Z) -2\frac{5+4r_Z}{1+2r_Z}\log(1-r_Z)\label{Ga}\\
&-&\frac{4r_Z(1+r_Z)(1-2r_Z)}{(1-r_Z)^2(1+2r_Z)}\log(r_Z)-
+\frac{5+9r_Z-6r_Z^2}{(1-r_Z)(1+2 r_Z)}-8\mathrm{Li}_2(r_Z) -\frac{4\pi^2}{3}
\bigg]\Bigg]\,,\nonumber\\
\Gamma_{b}^{Z} &=& \Gamma_{b}^{Z(0)}\Bigg[1+\frac{\alpha_s}{4\pi}C_F
\bigg[2\log\left(\frac{m_t^2}{\mu^2}\right) - 4 \log(1-r_Z)\log(r_Z)-\frac{2(8+r_Z)}{2+r_Z}\log(1-r_Z)\label{Gb}\\
&-&\frac{4r_Z(2-2r_Z-r_Z^2)}{(1-r_Z)^2(2+r_Z)}\log(r_Z)
-8\mathrm{Li}_2(r_Z)-\frac{16-11r_Z-17r_Z^2}{3(1-r_Z)(2+r_Z)} + 8 -\frac{4\pi^2}{3}\bigg]\Bigg]\,,\nonumber\\
\Gamma_{ab}^{Z} &=& \Gamma_{ab}^{Z(0)}\Bigg[1+\frac{\alpha_s}{4\pi}C_F
\bigg[
\log\left(\frac{m_t^2}{\mu^2}\right)-4\log(1-r_Z)\log(r_Z) -\frac{2(2+7r_Z)}{3r_Z}\log(1-r_Z)\label{Gab}\\
&-&\frac{4r_Z(3-2r_Z)}{3(1-r_Z)^2}\log(r_Z)+
 \frac{5-9r_Z}{3(1-r_Z)}+4-8\mathrm{Li}_2(r_Z)-\frac{4\pi^2}{3}\bigg]\Bigg]\,.\nonumber
\ee
\end{subequations}
Again we were able to crosscheck our results (\ref{GjustZ}) with the corresponding calculation done for a virtual photon contributing to the $B\to X_s l^+ l^-$ spectrum~\cite{Asatryan:2002iy}. 
After taking into account the different dipole operator renormalization condition in~\cite{Asatryan:2002iy} (including mass renormalization) we find complete agreement with their results. $\Gamma_a^{Z}$ was also cross-checked with the corresponding calculation of the $t\to W b$ decay width at NLO in QCD~\cite{Li:1990qf}. Finally, we have compared our $\Gamma^b_Z$ expression with the results given by Zhang et al.\ in ref.~\cite{Zhang:2008yn}. In the limit $r_Z \to 0$ our results agree with those given in \cite{Zhang:2008yn}, but we find disagreement in the $r_Z$ dependence. After our first publication of these results in~\cite{Drobnak:2010wh}, we were made aware of a new paper in preparation by the same authors, which has now been published \cite{Zhang:2010bm} and therein a corrected result for $\Gamma^b_Z$ is given that coincides with ours. 

The remaining bremsstrahlung contributions are induced by the gluon dipole operator and are IR finite
\begin{subequations}\label{XXX}
\be
\Gamma_{ag}^{Z,\mathrm{brems.}}&=&\frac{\Gamma^{Z(0)}_{ab}}{3(1-r_Z)^2} \frac{\alpha_s}{4\pi}C_F\Bigg[
2\hat v\bigg\llbracket \frac{1}{4}(3 -4 r_Z+r_Z^2)+\log(r_Z)(1-r_Z-r_Z^2)-\mathrm{Li}_2(1-r_Z)\label{Bremsag}\\
&+&r_Z\sqrt{(4-r_Z)r_Z}\Big(\arctan\Big(\sqrt{\frac{r_Z}{4-r_Z}}\Big)+\arctan\Big(\frac{r_Z-2}{\sqrt{(4-r_Z) r_Z}}\Big)\Big)\nonumber \\
&+&2\mathrm{Re}\Big\{\mathrm{Li}_2\Big(\frac{1}{2}(1-r_Z)(2-r_Z-\ii\sqrt{(4-r_Z)r_Z}\Big)\Big\}
\bigg\rrbracket \nonumber\\
&+&2\hat a\bigg\llbracket \frac{1}{4}(3 -8 r_Z+5r_Z^2)+\frac{1}{2}\log(r_Z)(-2-7r_Z+2r_Z^2)+\mathrm{Li}_2(1-r_Z)\nonumber\\
&+&(3-r_Z)\sqrt{(4-r_Z)r_Z}\Big(\arctan\Big(\sqrt{\frac{r_Z}{4-r_Z}}\Big)+\arctan\Big(\frac{r_Z-2}{\sqrt{(4-r_Z) r_Z}}\Big)\Big)\nonumber \\
&-&2\mathrm{Re}\Big\{\mathrm{Li}_2\Big(\frac{1}{2}(1-r_Z)(2-r_Z-\ii\sqrt{(4-r_Z)r_Z}\Big)\Big\}
\bigg\rrbracket\Bigg]\,,\nonumber
\ee
\be
\Gamma_{bg}^{Z,\mathrm{brems.}}&=& \frac{\Gamma_b^{Z(0)}}{(1-r_Z)^2 2(2+r_Z)} \frac{\alpha_s}{4\pi}C_F\Bigg[
2 \hat v\bigg\llbracket \frac{1}{3}(1-r_Z)(-25+2r_Z-r_Z^2)-4r_Z\log(r_Z)(1+r_Z)\label{Bremsbg}\\
&-&4(1-r_Z)\sqrt{(4-r_Z)r_Z}\Big(\arctan\Big(\sqrt{\frac{r_Z}{4-r_Z}}\Big)+\arctan\Big(\frac{r_Z-2}{\sqrt{(4-r_Z) r_Z}}\Big)\Big)-4\mathrm{Li}_2(1-r_Z)\nonumber \\
&+&8 \mathrm{Re}\Big\{\mathrm{Li}_2\Big(\frac{1}{2}(1-r_Z)(2-r_Z-\ii\sqrt{(4-r_Z)r_Z}\Big)\Big\}
\bigg\rrbracket \nonumber\\
&+&2\hat a\bigg\llbracket 9-r_Z(2+7r_Z)+r_Z\log(r_Z)(8+5r_Z)+4\mathrm{Li}_2(1-r_Z) \nonumber\\
&+&2(2-r_Z)\sqrt{(4-r_Z)r_Z}\Big(\arctan\Big(\sqrt{\frac{r_Z}{4-r_Z}}\Big)+\arctan\Big(\frac{r_Z-2}{\sqrt{(4-r_Z) r_Z}}\Big)\Big) \nonumber \\
&-&8 \mathrm{Re}\Big\{\mathrm{Li}_2\Big(\frac{1}{2}(1-r_Z)(2-r_Z-\ii\sqrt{(4-r_Z)r_Z}\Big)\Big\}
\bigg\rrbracket\Bigg]\,,\nonumber\\
\Gamma_{g}^{Z} &=& \frac{\Gamma_b^{Z(0)}}{(1-r_Z)^2 2(2+r_Z)}\frac{\alpha_s}{4\pi}C_F\Bigg[
\frac{\hat v^2}{6} \bigg\llbracket(1-r_Z)(77-r_Z-4 r_Z^2)+ 3\log(r_Z)(10-4r_Z-9r_Z^2) \label{Bremsg}\\
&+&6 \sqrt{\frac{r_Z}{4-r_Z}}(20+10r_Z-3r_Z^2)
   \Big(\arctan\Big(\sqrt{\frac{r_Z}{4-r_Z}}\Big)+\arctan\Big(\frac{r_Z-2}{\sqrt{(4-r_Z) r_Z}}\Big)\Big)
+12\log^2(r_Z)\nonumber \\
&+&48 \mathrm{Re}\Big\{\mathrm{Li}_2\Big(\frac{1}{2}+\frac{\ii}{2} 
   \sqrt{\frac{4-r_Z}{r_Z}}\Big)-\mathrm{Li}_2\Big(\frac{r_Z}{2}+\frac{\ii}{2} \sqrt{(4-r_Z) r_Z}\Big)\Big\}\bigg\rrbracket \nonumber\\
&+&\frac{\hat a^2}{6}\bigg\llbracket\frac{(1-r_Z)}{r_Z}(1-70r_Z+38r_Z^2-5r_Z^3)+3\log(r_Z)(2+46r_Z-9r_Z^2+4\log(r_Z)) \nonumber\\
&-&6(20-3r_Z)\sqrt{(4-r_Z)r_Z}\Big(\arctan\Big(\sqrt{\frac{r_Z}{4-r_Z}}\Big)+\arctan\Big(\frac{r_Z-2}{\sqrt{(4-r_Z) r_Z}}\Big)\Big)\nonumber \\
&+&48 \mathrm{Re}\Big\{\mathrm{Li}_2\Big(\frac{1}{2}+\frac{\ii}{2} 
   \sqrt{\frac{4-r_Z}{r_Z}}\Big)-\mathrm{Li}_2\Big(\frac{r_Z}{2}+\frac{\ii}{2} \sqrt{(4-r_Z) r_Z}\Big)\Big\}\bigg\rrbracket\nonumber \\
&+&\hat a \hat v \bigg\llbracket - 7 + 22 r_Z -15 r_Z^2 -\log(r_Z)(6-5r_Z^2+4\log(r_Z))\nonumber\\
&+&2(2+r_Z)\sqrt{(4-r_Z)r_Z}\Big(\arctan\Big(\sqrt{\frac{r_Z}{4-r_Z}}\Big)+\arctan\Big(\frac{r_Z-2}{\sqrt{(4-r_Z) r_Z}}\Big)\Big)\nonumber \\
&-&16 \mathrm{Re}\Big\{\mathrm{Li}_2\Big(\frac{1}{2}+\frac{\ii}{2} 
   \sqrt{\frac{4-r_Z}{r_Z}}\Big)-\mathrm{Li}_2\Big(\frac{r_Z}{2}+\frac{\ii}{2} \sqrt{(4-r_Z) r_Z}\Big)\Big\}\bigg\rrbracket \Bigg]\,.\nonumber
\ee
\end{subequations}
Numerical agreement was found when comparing the vector parts of our results given in eq.~(\ref{XXX}) with the corresponding results of ref. \cite{Asatryan:2002iy}.

\subsection{Numerical analysis}
In this section we present some numerical values to estimate the significance of QCD corrections. In particular we parametrize the decay width given in eq.~(\ref{dw}) as
\be
\Gamma&=&\frac{m_t}{16\pi}g_Z^2\Bigg\{
\frac{v^4}{\Lambda^4}|a^Z|^2  \Big[x_{a} +\frac{\alpha_s}{4\pi}C_F y_{a} \Big]+
\frac{v^2m_t^2}{\Lambda^4}|b^Z|^2  \Big[ x_{b}+\frac{\alpha_s}{4\pi}C_F y_{b}\Big]
+\frac{v^3 m_t}{\Lambda^4}2\mathrm{Re}\{b^{Z*} a^Z\}\Big[x_{ab} +\frac{\alpha_s}{4\pi}C_F y_{ab}\Big]\\
&+&|b^g|^2 \frac{v^2m_t^2}{\Lambda^4}\frac{\alpha_s}{4\pi}C_F y_{g}
+\frac{v^3 m_t}{\Lambda^4}\Big[
2\mathrm{Re}\{a^{Z*} b^{g}\}\frac{\alpha_s}{4\pi}C_F y_{ag}
-2\mathrm{Im}\{a^{Z*} b^{g}\}\frac{\alpha_s}{4\pi}C_F \tilde{y}_{ag}\Big] \nonumber\\
&+&\frac{v^2m_t^2}{\Lambda^4}\Big[2\mathrm{Re}\{b^{Z*}b^g\}\frac{\alpha_s}{4\pi}C_F y_{bg}-2\mathrm{Im}\{b^{Z*}b^g\}\frac{\alpha_s}{4\pi}C_F \tilde{y}_{bg} \Big] \Bigg\}\nonumber\,.
\ee
Here $x_i$ stand for the tree-level contributions, while $y_i\,, \tilde{y}_i$ denote the corresponding QCD corrections. Numerical values of the coefficients are given in Table \ref{table:num}. We see that corrections due to the gluon dipole operator are an order of magnitude smaller (except $y_g$, which is even more suppressed) than corrections to the $Z$ operators themselves and have opposite sign. 
\begin{table}[h]
\begin{center}
\begin{tabular}{llll}
\hline\hline
$x_{b}=2.36$ & $x_{a}=1.44$ & $x_{ab}=1.55$   \\
$y_{b}=-17.90$ & $y_{a}=-10.68$ & $y_{ab}=-10.52$ &$y_{g}=0.0103$\\
 $y_{bg}=3.41$ & $y_{ag}=2.80$ & $\tilde{y}_{bg}=2.29$ &  $\tilde{y}_{ag} = 1.50$\\
 \hline\hline
\end{tabular}
\caption{\label{table:num}Numerical values of coefficient functions at the renormalization scale equal to the top quark mass corresponding to the inputs
$m_t = 172.3\,\mathrm{GeV}\,,$  $m_Z = 91.2\,\mathrm{GeV}\,,$  $\sin^2\theta_W = 0.231\,.$}
\end{center}
\end{table}

Next we investigate the relative change of the decay rates and branching ratios when going from the leading order to next to leading order in QCD, where branching ratio for the top quark is defined to be normalized to its main decay channel $\mathrm{Br}(t\to qZ,\gamma) = \Gamma(t\to q Z,\gamma)/\Gamma(t\to b W)$.
\begin{table}[!h]
\begin{center}
\begin{tabular}{l|l|l|l||l|l}
	&$b^Z=b^g=0$&$a^Z=b^g=0$&$a^Z=b^Z, b^g=0$& $b^Z=0, a^Z =b^g$ &$a^Z=0, b^Z =b^g$\\
\hline\hline
$\Gamma^{\mathrm{NLO}}/\Gamma^{\mathrm{LO}}$&$0.92$& $0.91$& $0.92$ & $0.95$&$0.94$\\
$\Brfrac$&$1.001$&$0.999$&$1.003$ &$1.032$&$1.022$\\
\hline\hline
\end{tabular}
\caption{Numerical values of $\Gamma^{\mathrm{NLO}}/\Gamma^{\mathrm{LO}}$ and $\mathrm{Br}^{\mathrm{NLO}}(t\to q Z)/\mathrm{Br}^{\mathrm{LO}}(t\to q Z)$ for certain values and relations between Wilson coefficients. Results are obtained using analytical formulae given above setting $\mu=m_t$. Main decay channel decay width at NLO in QCD is given in ref. \cite{Li:1990qf}. Additional SM inputs used are $m_W=80.4\,\mathrm{GeV}$ and $\alpha_s=0.107$.}
\label{brsZ}
\end{center}
\end{table}
The results are presented in Table \ref{brsZ}. We see that the change in the decay width is of the order 10\%. There is a severe cancellation between the QCD corrections to $\Gamma(t\to c Z)$ and the main decay channel $\Gamma(t\to b W)$. This cancellation causes the change of the branching ratio to be only at the per-mile level when $b^g$ is set to zero. In the case when only operators $\mathcal O_{L,R}^Z$ are considered this cancellation is anticipated since the NLO correction to $\Gamma^Z_{a}$ is of the same form as the correction to the rate of the main decay channel. If we treat $b$ quarks as massless, exact cancellation is avoided only due to the difference in the masses of $Z$ and $W$ bosons. It is more surprising that similar cancellation is obtained also when only the dipole $Z$ operator is considered. However, setting $b^g=a^Z$ or $b^g=b^Z$, the impact of QCD corrections is increased by an order of magnitude and reaches a few percent.

\section{$t\to q \gamma$ decay}
Compared to eq.~(\ref{dw}), the expression for the total $t\to q \gamma$ decay width gets simplified, since due to gauge invariance we only need to consider dipole operators
\be
\Gamma^{\gamma} &=& \frac{v^2 m_t^2}{\Lambda^4}\left[ |b^{\gamma}|^2 \Gamma^{\gamma}_{b} 
+|b^g|^2 \Gamma^{\gamma}_{g} +2\mathrm{Re}\{b^{\gamma*}b^g\} \Gamma^{\gamma}_{bg} -2\mathrm{Im}\{b^{\gamma*}b^g\}\tilde{\Gamma}^{\gamma}_{bg} \right]\,.
\ee
We follow a similar pattern of analysis as in the $Z$ case, with the added complication that the bremsstrahlung contributions have to be treated in greater detail since the non-trivial photon spectrum has important implications for the experimental detection of this decay channel. We start with the tree level contribution, which we write in $4+\epsilon$ dimensions as
\be
\Gamma^{\gamma(0)}_b = \lim_{\epsilon \to 0}m_t \alpha (1+\frac{\epsilon}{2})\Gamma(1+\frac{\epsilon}{2})\,.
\ee

\subsection{Virtual corrections}
Feynman diagrams yielding virtual corrections to the $t\to q\gamma$ decay rate are presented in Figure \ref{virtcorrections}. We are dealing with one less diagram, since we only have the photonic dipole operator. Quark field renormalization remains the same and the decay amplitude can be written in terms of two form factors
\be
A_{t\to q\gamma} &=& \frac{v}{\Lambda^2}\Big[ b^{\gamma} \big(1+\frac{\alpha_s}{4\pi}C_F F^{\gamma}_b\big)
  +  b^g \frac{\alpha_s}{4\pi}C_F F_{bg}^{\gamma}\Big]\langle \op_{LR,RL}^{\gamma}\rangle\,,
\ee
which read
\be
F_{b}^{\gamma}&=&C_{\epsilon}
\Bigg[-\frac{4}{\epsIR^2}+\frac{5}{\epsIR}+\frac{2}{\epsUV} -6 \Bigg] + \delta_{77}^{\gamma}\,,\\
F_{bg}^{\gamma}&=&Q C_{\epsilon}
\Bigg[\frac{8}{\epsUV}-11+\frac{2}{3}\pi^2-2\pi\ii \Bigg]+ \delta_{87}^{\gamma}\,.
\ee
Operator renormalization counter terms induced by the UV matching procedure in the $\overline{\mathrm{MS}}$ scheme read
\be
Z_{b}^{\gamma}  &=& 1+\frac{\alpha_s}{4\pi}C_F\delta_{b}^{\gamma}\,,\hspace{0.5cm}
\delta_{b}^{\gamma}= - \Big(\frac{2}{\epsUV}\ +\gamma-\log(4\pi)\Big)\,,\label{renF}\\
Z_{bg}^{\gamma} &=& 1+\frac{\alpha_s}{4\pi}C_F\delta_{bg}^{\gamma}\,,\hspace{0.5cm}
\delta_{bg}^{\gamma}= -4 Q \Big(\frac{2}{\epsUV} +\gamma-\log(4\pi)\Big)\,.\nonumber
\ee
Finally, tree level and one-loop virtual correction contributions to the $t\to q \gamma$ decay width are
\begin{subequations}
\be
\Gamma_{b}^{\gamma,\mathrm{virt.}}&=&\Gamma_{b}^{\gamma(0)}
\bigg[1+\frac{\alpha_s}{4\pi}C_F\Big[-\frac{8}{\epsIR^2}+\frac{6}{\epsIR}-7-\frac{\pi^2}{3}+2\log\left(\frac{m_t^2}{\mu^2}\right)\Big]\bigg]\,,\label{GbvF}\\
\Gamma_{bg}^{\gamma,\mathrm{virt.}}&=&\Gamma_{b}^{\gamma(0)}
\frac{\alpha_s}{4\pi}C_F Q\Big[-11+\frac{2\pi^2}{3}+4\log\left(\frac{m_t^2}{\mu^2}\right)\Big]\,,\\
\tilde{\Gamma}_{bg}^{\gamma,\mathrm{virt.}}&=&\Gamma_{b}^{\gamma(0)}
\frac{\alpha_s}{4\pi}C_F Q\Big[-2\pi\Big]\,.
\ee
\end{subequations}
As in the $t\to q Z$ case we are left with IR divergences which have to be canceled by the corresponding bremsstrahlung contributions.

\subsection{Bremsstrahlung corrections}
\begin{figure}[t]
\begin{center}
\includegraphics[width=7.0cm]{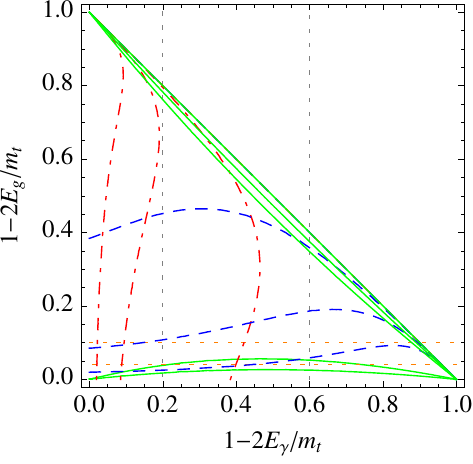}
\end{center}
\caption{ \label{fig:dalitz} The $t\to q \gamma g$ Dalitz plot. Contours of constant photon and gluon infrared and collinear divergent contributions are drawn in red (dot-dashed) and blue (dashed) lines respectively. The collinear divergencies appear at the horizontal and vertical boundaries of the phase-space, while the IR divergencies sit in the top and right corners. The cuts on the photon energy correspond to vertical lines, the cuts on the gluonic jet energy to horizontal lines. Full green lines correspond to cuts on the jet veto cone around the photon. }
\end{figure}
The $t\to q \gamma g$ decay process involves three (one almost) massless particles in the final state. Virtual matrix element corrections contribute only at the soft gluon endpoint ($E_g = 0$) and result in non-vanishing  $b^\gamma b^g$ interference contributions. They involve IR divergencies which are in term canceled by the real gluon emission contributions. These also produce non-vanishing $|b^g|^2$ contributions, and  create a non-trivial photon spectrum involving both soft and collinear divergencies. The later appear whenever a photon or a gluon is emitted collinear to the light quark jet.  An analogous situation is encountered in the $B\to X_s\gamma$ decay measured at the $B$-factories. However, there the photon energy in the $B$ meson frame can be reconstructed and a hard cut ($E_\gamma^{\mathrm{cut}}$) on it removes the soft photon divergence. The cut also ensures that the $B\to X_s g$ process contributing at the end-point $E_\gamma = 0$ is suppressed. On the other hand, in present calculations the collinear divergencies are simply regulated by a non-zero strange quark mass, resulting in a moderate $\log(m_s/m_b)$ contributions to the rate.  The situation at the Tevatron and the LHC is considerably different.  The initial top quark boost is not known and the reconstruction of the decay is based on triggering on isolated hard photons with a very loose cut on the photon energy (a typical value being $E_\gamma>10 $\,GeV in the lab frame \cite{Aad:2009wy}). Isolation criteria are usually specified in terms of a jet veto cone $\Delta R = \sqrt {\Delta \eta^2 + \Delta \phi^2}$ where $\Delta\eta$ is the difference in pseudorapidity and $\Delta\phi$ the difference in azimuthal angle between the photon and nearest charged track. Typical values are $\Delta R > (0.2-0.4)$ \cite{CDF_2007_06}. We model the non-trivial cut in the top quark frame with a cut on the projection of the photon direction onto the direction of any of the two jets ($\delta r= 1- {\bf p}_\gamma \cdot {\bf p}_j / E_\gamma E_j$). The effects of the different cuts on the decay Dalitz plot are shown in Figure \ref{fig:dalitz}.
Since at this order there are no photon collinear divergencies associated with the gluon jet, the $\hat{x}$ cut around the gluon jet has a numerically negligible effect on the rate. On the other hand the corresponding cut on the charm jet - photon separation does not completely remove the divergencies in the spectrum. However, they become integrable. The combined effect is that the contribution due to the gluonic dipole operator can be enhanced compared to the case of $B\to X_s \gamma$. Below we give the full analytical formulae for the $t\to q \gamma g$ decay rate including the effects of kinematical cuts on variables
$\hat{x} \equiv \delta r$ and 
$\hat{y} \equiv 2 E_{\gamma}^{\mathrm{cut}}/m_t$.
\begin{subequations}
\be
\Gamma_{b}^{\gamma,\mathrm{brems.}}&=&\Gamma_{b}^{\gamma(0)} \frac{\alpha_s}{4\pi}C_F\Bigg[
\frac{8}{\epsIR^2}-\frac{6}{\epsIR}+1-\pi^2-2\frac{\uc(1-\uc)(2\uc-1)}{2-\uc \xc}+\uc 
+\frac{4}{\xc}(2-\uc)(1-\uc)-16\frac{1-\uc}{\xc^2}\label{GbbF}\\
&-&2\log^2(1-\uc)+(\uc^2+2\uc-10)\log(1-\uc)-\frac{2\xc^2-24 \xc+32}{\xc^3}\log\Big(\frac{2-\xc}{2-\uc\xc}\Big)\nonumber\\
&-&6\log\Big(\frac{2-\xc}{2-\xc\uc(2-\uc)}\Big)-\Big(\frac{2}{\xc}+\uc^2+2\uc\Big)
\log\Big(\frac{2-\xc\uc(2-\uc)}{2-\uc\xc}\Big)\nonumber\\
&+&12\sqrt{2/\xc-1}\arctan\Big(\frac{1-\uc}{\sqrt{2/\xc-1}}\Big)
+4\mathrm{Li}_2\Big(\xc \frac{1-\uc}{\xc -2}\Big)-2\mathrm{Li}_2\Big(\xc\frac{(1-\uc)^2}{\xc-2}\Big)
\Bigg]\,,\nonumber\\
\Gamma_{bg}^{\gamma,\mathrm{brems.}}&=&
\Gamma_{b}^{\gamma(0)} \frac{\alpha_s}{4\pi}C_F Q\Bigg[
-\frac{(1-\uc)(2-\xc)(\uc\xc^2-2\uc\xc-2\xc+8)}{\xc^2(2-\uc\xc)}+\frac{2\pi^2}{3}-
4(1-\uc)\log\Big(\frac{2-\xc\uc(2-\uc)}{(1-\uc)(2-\uc\xc)}\Big)\\
&-&4\log(\uc)\log\Big(\frac{2-\xc\uc(2-\uc)}{2}\Big)+2\log\Big(\frac{\xc}{2}\Big)\log\Big(\frac{2-\xc}{2-\xc\uc(2-\uc)}\Big)\nonumber \\ 
&-&\frac{4}{\xc^3}(\xc^2-4\xc+4)\log\Big(\frac{2-\xc}{2-\xc\uc}\Big)
+4\Big(\mathrm{Li}_2\Big(\frac{\xc}{2}\Big)-\mathrm{Li}_2(\uc)-\mathrm{Li}_2\Big(\frac{\xc\uc}{2}\Big)\Big)\nonumber\\
&-&8\arctan\Big(\frac{1-\uc}{\sqrt{2/\xc-1}}\Big)\Big(\sqrt{2/\xc-1}- \arctan(\sqrt{2/\xc-1}\Big) \nonumber\\
&+&8\mathrm{Re}\Big\{\mathrm{Li}\Big(\frac{1}{2}\big(2-\xc-\ii\sqrt{(2-\xc)\xc}\big)\Big)-
\mathrm{Li}_2\Big(\frac{1}{2}\big(2-\xc\uc-\ii \uc\sqrt{(2-\xc)\xc}\big)\Big)\Big\}\Bigg] \nonumber\,,\\
\Gamma_{g}^{\gamma}&=&
\Gamma_{b}^{\gamma(0)} \frac{\alpha_s}{4\pi}C_F Q^2\Bigg[-\frac{(1-\uc)(2-\xc)(3\uc\xc^2-4\xc\uc-8\xc+16)}{\xc^2(2-\xc\uc)}+\frac{2\pi^2}{3}+
\big(4-2\xc+4\log\Big(\frac{\xc}{2}\Big)\big)\log(\uc)\\
&+&(3-\uc)(1-\uc)\log\Big(\xc\frac{1-\uc}{2-\xc \uc}\Big)
+\frac{2}{\xc^3}(2-\xc)(\xc^3-\xc^2+6\xc-8)\log\Big(\frac{2-\xc}{2-\xc\uc}\Big)\nonumber\\
&+&4\Big(\mathrm{Li}_2\Big(\frac{\xc\uc}{2}\Big)-\mathrm{Li}_2\Big(\frac{\xc}{2}\Big)-\mathrm{Li}_2(\uc)\Big)
\Bigg]\,.\nonumber
\ee
\end{subequations}
It is easy to verify the cancellation of IR divergences upon summation of $\Gamma_{b}^{\gamma,\mathrm{virt.}}$ and $\Gamma_{b}^{\gamma,\mathrm{brems.}}$ given in eq.~(\ref{GbvF}) and (\ref{GbbF}) respectively.

\subsection{Numerical analysis}
\begin{figure}[t]
\begin{center}
\includegraphics[width=7.0cm]{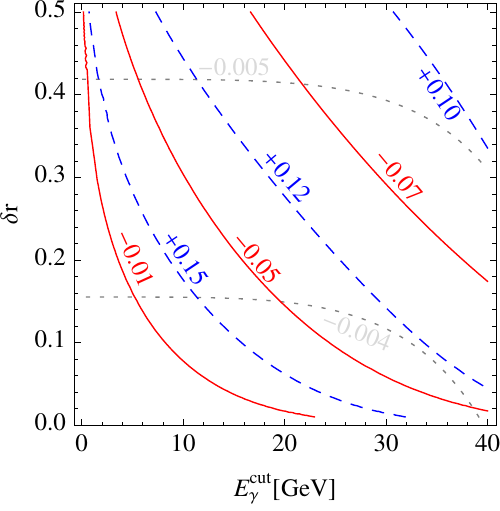}
\end{center}
\caption{ \label{fig:drEcut}  Relative size of $\alpha_s$ corrections to the $\mathrm{Br}(t\to q \gamma)$ at representative ranges of $\delta r$ and $E^{\mathrm{cut}}_\gamma$. Contours of constant correction values are plotted for $b^g=0$ (gray, dotted), $b^g=b^\gamma$ (red) and $b^g = - b^\gamma$ (blue, dashed).}
\end{figure}
\begin{figure}[t]
\begin{center}
\includegraphics[width=10.0cm]{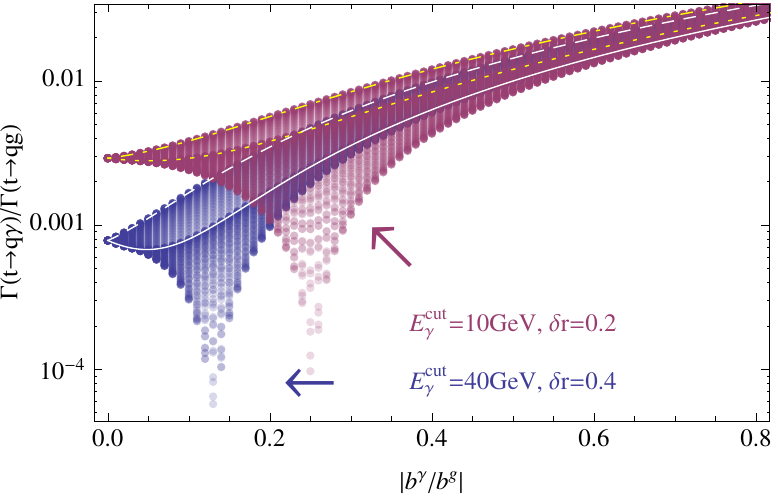}
\end{center}
\caption{ \label{fig:tcg-tcg}  The ratio of radiative rates $\Gamma(t\to q \gamma) / \Gamma(t\to q g )$ versus the absolute ratio of the relevant effective FCNC couplings $|b^\gamma/b^g|$. Two representative choices of experimental kinematic cuts are shown. The shaded bands represent the possible spread due to the unknown relative phase between $b^\gamma$ and $b^g$ couplings, while the lines correspond to maximal positive (full, dotted) and negative (dashed, dot-dashed) interference $b^\gamma b^g$. See text for details.}
\end{figure}

In Figure \ref{fig:drEcut} we show the $b^g$ induced correction to the tree-level $\mathrm{Br}(t\to q \gamma)$ for representative ranges of $\delta r$ and $E^{\mathrm{cut}}_\gamma$.
We observe, that the contribution of $b^g$ can be of the order of $10-15\%$ of the total measured rate, depending on the relative sizes and phases of $\mathcal O_{LR,RL}^{g,\gamma}$ and on the particular experimental cuts employed. Consequently, a bound on $\mathrm{Br}(t\to q \gamma)$ can, depending on the experimental cuts, probe both $b^{g,\gamma}$ couplings. In order to illustrate our point, we plot  the ratio of radiative rates $\Gamma(t\to q \gamma) / \Gamma(t\to q g )$, both computed at NLO in QCD versus the ratio of the relevant effective FCNC dipole couplings $|b^\gamma/b^g|$ in figure~\ref{fig:tcg-tcg}. 
We show the correlation for two representative choices of experimental kinematic cuts for the $t\to q \gamma$ decay. The vertical spread of the bands is due to the variation of the relative phase between $b^\gamma$ and $b^g$ couplings. We also display the two interesting limits where the $b^\gamma b^g$ interference is maximal positive (zero relative phase) and negative (relative phase $\pi$). We see that apart from the narrow region around $|b^\gamma/b^g| \sim 0.2$, where the two contributions may be fine-tuned and conspire to diminish the total $t\to c \gamma$ rate, the two radiative rates are well correlated. In particular, depending on the kinematical cuts employed, there is a natural lower bound on ratio of decay rates, valid outside of the fine-tuned region. Finally, for  $|b^\gamma/b^g|>0.6$ the correlation becomes practically insensitive to the particular experimental cuts employed and also the unknown relative phase between $b^\gamma$ and $b^g$ couplings. 

\section{Conclusions}
In summary, QCD corrections to FCNC coupling mediated rare top decays can induce sizable mixing of the relevant operators, both through their renormalization scale running~\cite{Drobnak:2010wh} as well in the form of finite matrix element corrections. These effects are found to be relatively small for the $t\to c Z$ decay. On the other hand the accurate interpretation of experimental bounds on radiative top processes in terms of effective FCNC operators requires the knowledge of the experimental cuts involved and can be used to probe $\mathcal O^g_{LR,RL}$ contributions indirectly. 

Finally we note that additional information on the underlying NP contributions is provided by kinematical distributions of $t\to c Z,\gamma$ common final states such as charged di-leptons~\cite{Drobnak:2008br} as well as in the case of $t\to c g$ through single top production cross-section~\cite{Malkawi:1995dm,Hosch:1997gz,Han:1998tp}. Combined, these observables could facilitate the reconstruction of  prospective NP models in  case a positive experimental signal of FCNC top quark processes would emerge in the future.

\begin{acknowledgments}
J. F. K. would like to thank Mikolaj Misiak and Gino Isidori for useful discussions and the Galilo Galilei Institute for Theoretical Physics for the hospitality and the INFN for partial support during the completion of this work. This work is supported in part by the European Commission RTN network, Contract No. MRTN-CT-2006-035482 (FLAVIAnet) and by the Slovenian Research Agency.
\end{acknowledgments}

\bibliography{reference}

\end{document}